\documentclass[numsec,webpdf,modern,small]{oup-authoring-template}

\onecolumn 

\usepackage{amssymb}
\usepackage{siunitx}
\usepackage{amsmath}
\usepackage{float}
\usepackage{gensymb}
\usepackage{upgreek}
\graphicspath{{Fig/}}

\usepackage[right]{lineno}
\usepackage{setspace}

\begin{document}

\journaltitle{}
\DOI{}
\copyrightyear{2024}
\pubyear{2019}
\access{Advance Access Publication Date: Day Month Year}
\appnotes{Paper}

\firstpage{1}


\title[cNMF for EBSD]{Employing constrained non-negative matrix factorization for microstructure segmentation}

\author[1,$\ast$]{Ashish Chauniyal}
\author[2]{Pascal Thome}
\author[1]{Markus Stricker}

\authormark{Ashish Chauniyal et al.}

\address[1]{\orgdiv{Interdisciplinary Centre for Advanced Materials Simulation (ICAMS)}, \orgname{Ruhr-Universit\"at Bochum}, \orgaddress{\street{Univeristätstraße 150}, \postcode{44780},  \country{Germany}}}
\address[2]{\orgdiv{Institute for Materials (IFM)}, \orgname{Ruhr-Universit\"at Bochum}, \orgaddress{\street{Univeristätstraße 150}, \postcode{44780}, \country{Germany}}}

\corresp[$\ast$]{Corresponding author. \href{email:ashish.chauniyal@rub.de}{ashish.chauniyal@rub.de}}

\received{Date}{0}{Year}
\revised{Date}{0}{Year}
\accepted{Date}{0}{Year}



\abstract{Materials characterization using electron backscatter diffraction (EBSD) requires indexing the orientation of the measured region from Kikuchi patterns. The quality of Kikuchi patterns can degrade due to pattern overlaps arising from two or more orientations, in the presence of defects or grain boundaries. In this work we employ constrained non-negative matrix factorization to segment a microstructure with small grain misorientations,~\mbox{($<1\degree$)}, and predict the amount of pattern overlap. First we implement the method on mixed simulated patterns - that replicates a pattern overlap scenario, and demonstrate the resolution limit of pattern mixing or factorization resolution using a weight metric. Subsequently, we segment a single-crystal dendritic microstructure and compare the results with high resolution EBSD. By utilizing weight metrics across a low angle grain boundary we demonstrate how very small misorientations/low-angle grain boundaries can be resolved at a pixel level. Our approach constitutes a versatile and robust tool, complementing other fast indexing methods for microstructure characterization.}

\maketitle

\section{Introduction}
\label{sec:Intro} 
Electron backscatter diffraction (EBSD) has been the prime workhorse for the analysis of crystal structures and orientation of crystalline materials for many years~\citep{EnglerandRandle2010,graef_book}. While the fundamental principles of diffraction and recording of Kikuchi patterns have not changed over the years, there has been continuous developments in novel ways of post-processing. Starting from Hough-based indexing~\citep{EBSD1,EBSD_Hough,Hough,Lassen}, which transforms the patterns to a Hough space followed by indexation, to the more recent Dictionary Indexing (DI)~\citep{HREBSD_graef, HREBSD_patternmatchingTanaka,Nolze,Nolze2} and spherical indexing (SI)~\citep{Overlap_Lenthe,sphericalharmonics_Lenthe2} approaches, which directly compare the patterns to a simulated library and find out the best match using cross-correlation. Such indexing methods can effectively process noisy patterns and also provide good lateral and angular resolutions. The angular resolutions of EBSD has further been improved using cross-correlation methods that measure pattern shifts, along with the ability to quantify strains~\citep{HREBSD_Wright_strain,Wilkinson, HREBSD_Wilkinson, HREBSD_Britton, RVB_EBSD, HREBSD_wilkinson2, HREBSD_patternmatchingTanaka}. In addition to these cross-correlation methods, the angular resolution can further be improved using pattern matching based refinement approach ~\citep{EBSD_Nolze,patternmatching_refine1, patternmatching_refine2,patternmatching_refine3} and DIC type approaches ~\citep{DIC_1,DIC_2,DIC_3}. Overall, such methods provide a high resolution of angular variations by directly identifying sub-pixel level pattern shifts and go by the acronym high-resolution electron backscattered diffraction (HR-EBSD). 

Onwards from here, with the progress of machine learning in material science, several data-based supervised machine learning (ML) methods have recently been employed for indexation and segmentation of microstructures~\citep{Jha2018, CNN_Shen, Kaufmann_phasemapping, Kaufmann_MLdiffraction, CNN_graef, Deeplearning_Qi}. These methods circumvent the need to evaluate computationally expensive pattern comparisons and instead train a ML model that is adapted to the intricacies of the orientation space. While for routine tasks such methods are well suited, their robustness remains in question as machine conditions, sample preparations, crystallography, and the need for large training datasets can greatly limit the extent to which such methods can be deployed quickly. An additional aspect is their transferability to \textit{unseen} data. Due to the above reasons, it is generally desirable to employ unsupervised ML methods. 

Unsupervised methods promise a higher efficiency and robustness. Several such unsupervised data-based approaches have been presented~\citep{McAuliffe, Brewer_Multivariate, multivariate} which can discriminate between similar or dissimilar phases and improve the effective spatial resolution of automated EBSD analysis, particularly segmentation tasks. Denoising of electron backscattered diffraction patterns (EBSPs) is possible using (unsupervised) dimensionality reduction strategies~\citep{kikuchipy}. For unsupervised methods an essential precept is that EBSPs are not directly mapped to orientations or phases in the standard EBSD sense, but instead undergo dimensionality reduction on raw pattern data followed by clustering~\citep{multivariate,McAuliffe,Brewer_Multivariate}. Herein, EBSP patterns are treated as linear combinations of spatially simple components~\citep{Brewer_Multivariate}, which can then be used efficiently for segmentation. While the above methods do not directly provide the orientation information of the microstructure, they are still very useful in conjunction with standard EBSD methods. The most important advantage in this respect lies in the ability to segment and categorise data without prior knowledge of the crystallography of microstrucutre. This implies that these methods are easily transferable to different crystallographic phases and microstructures, without the need for expensive supervised training. Where the latter often involves human-created masks or labels because these methods detect pattern existing in data without human bias.

The simplest dimensionality reduction algorithms that are used for EBSD data analysis, are principle component analysis (PCA)~\citep{Brewer_Multivariate,multivariate,kikuchipy} and non-negative matrix factorization (NMF)~\citep{McAuliffe}.  NMF is shown to perform especially well for the segmentation of crystallographically similar phases, which extols the sensitivity of the technique in detecting very small changes in an EBSPs~\citep{McAuliffe}. So far, only a canonical NMF method has been implemented for segmentation which is essentially a subset of PCA. However, the completely unsupervised nature of NMF implies that the derived NMF coefficients may not be amenable to physical interpretation. The derived NMF coefficients look similar to an EBSP with only small variations in intensities~\citep{McAuliffe}, however from an EBSD standpoint they cannot be directly associated with an orientation or a phase. For a connection to physics, the coefficients in such a factorization should be related to the either an orientation or a phase, as well, which is not possible with canonical NMF. For this reason we need a factorization scheme that can exploit the prior knowledge of orientation or phases. Such a specialized factorization method is called constrained non-negative matrix factorization (cNMF) which incorporates the prior knowledge of one or all components during the factorization~\citep{cNMF}. In the context of EBSD, this translates to  having a prior knowledge of certain EBSPs belonging to a particular orientation or phase. These known EBSPs need to be supplied externally during factorization and therefore the scheme can be classified as semi-supervised. In this work we implement a cNMF algorithm to factorize EBSPs using known EBSPs and calculate the weights of the components. These weights thus provide a metric to estimate the fraction of known EBSPs in any pattern. We show how this approach can be used to segment microstructures in great detail, which is on par with HR-EBSD~\citep{RVB_EBSD} methods, with much less computational overhead.

Fixing components before factorization implies that we select certain EBSPs from the microstructure and then resolve the rest in terms of their weights with respect to these EBSPs. This approach of using \textit{weight-based} metric to resolve a microstructure, subsequently allows us to address the problem of pattern overlaps at grain boundaries or other defects~\citep{wright}. The subject of pattern overlaps has been explored extensively in the past~\citep{Overlap_Bate, Overlap_Zaefferer, wright, Overlap_Tong, Overlap_Singh, Overlap_Tripathi, Overlap_Lenthe, Overlap_Shi_GB, Overlap_BroduAimo, Overlap_GB_position} as it is well known that EBSD measurements nearby grain boundaries can create uncertainties in indexation. Inherent to EBSD measurements is the fact that the intensity contribution to the final pattern from each grain, when measured at a grain boundary, is related to the respective interaction volume of each grain. So far, the pattern overlap problem has been treated using rigorous pattern matching approaches which incorporate multiple crystal orientations~\citep{Overlap_Shi_GB,Overlap_Tong,Overlap_Lenthe}, or by determining a resolution limit to avoid overlaps in the first place~\citep{Overlap_Bate, Overlap_Zaefferer, Overlap_GB_position}. Yet, these approaches rely on mapping EBSPs to orientations. Even if these approaches lead to increase in accuracy, the volume fraction contribution from each grain may still remain unknown. Though the consideration of volume fraction has recently been made by simulating the mixing of Kikuchi patterns and subsequently matching patterns~\citep{Overlap_Lenthe}. Still, there is a need for employing a much less computationally expensive and more robust scheme.

In this regard, the work of ~\cite{Brewer_Multivariate} is most relevant to our work, as it employs a very simple yet robust PCA method to segment grains. Not only does this method deliver angular resolutions on par with standard EBSD, but can also discriminate between crystallographically similar phases. Furthermore, the authors foresaw the ability of PCA in deconvolving pattern overlaps, which can exceed the capabilities of EBSD. At the core of such analysis is the shift to a paradigm that EBSPs can be analyzed using an alternate metric, instead of orientations alone. In this work we follow this new paradigm of using alternate weight-based metric to characterize a microstructure~\citep{Brewer_Multivariate,multivariate,McAuliffe}. We employ an advanced data-based method - cNMF, which allows us to estimate the contribution of individual EBSPs in an overlapped EBSP using the weight metric. In this paper we briefly describe the cNMF method and explain its implementation for EBSPs. Thereafter, we first test the limits of the method on simulated patterns. We then present results pertaining to an experimental data set of low angle grain boundaries to showcase the segmentation ability for a limiting case of very small misorientations. This data set is specially chosen as it is synonymous to crystallographically similar EBPSs. Finally, we elaborate how to extract more information from the predicted values and discuss the identification of grain boundaries.

\section{Methods}
\subsection{Experimental Details}
\label{Methods:Experimental}
The specimen used for investigation is a cross-section of a cylindrical bar of single-crystalline superalloy ERBO1/C which is directionally solidified using a seeded Bridgman technique~\citep{Hallensleben}. This casting technique results in a single crystal along $\langle001\rangle$ direction. However, during solidification individual dendrites frequently deviate by small misorientation angles, which leads to a fine substructure~\citep{RVB_EBSD}. Cross-sections are carefully polished using a sequence of  $6 \mu$m, $3 \mu$m and $1\,\mu$m diamond suspensions, followed by a $3\,$h Vibrometer polishing with a colloidal silica suspension of $0.25\,\mu$m size. Finally, the surface has been cleaned by ion milling with $6\,$kV beam for $30\,$min at $40\degree$ inclination. The above steps are essential to obtain good quality Kikuchi patterns.

EBSD measurements are performed using a FEI Quanta 650 FEG SEM at $30\,$kV accelerating voltage, $60\,\mu$m aperture, $17\,$mm working distance WD and $200\,$mm camera length (distance between detector and sample). Patterns are recorded using an EDAX-TSL system with a Hikari-XP camera at a resolution of $512\times512$ pixels, with a step size of $1\,\mu$m.  

\subsection{Constrained Non-Negative Matrix Factorization}
\label{Methods:cNMF}
Constrained non-negative matrix factorization (cNMF)~\citep{cNMF} is essentially a version of the canonical non-negative matrix factorization (NMF) where the former is semi-supervised and the latter is unsupervised. Starting with NMF first, for a given non-negative matrix $\mathbf{X}$ of dimensions $m \times n$ we seek a factorization such that 
\begin{equation}
    \mathbf{X} \approx \mathbf{WH},
    \label{eq:equation1}
\end{equation}
where $\mathbf{W}$ and $\mathbf{H}$ are non-negative matrices of dimensions $m\times k$ and $k\times n$ respectively. Here $k$ is the number of components which are provided by the user. The above then reduces to a minimization problem which can be solved as 
\begin{equation}
   \min_{\mathbf{W}\in\mathbb{R}^{m\times k}, \mathbf{H}\in\mathbb{R}^{k\times n}}  D(\mathbf{X}, \mathbf{WH}),
\end{equation}
such that $\mathbf{W}\ge 0, \mathbf{H}\ge 0$ and $D$ is a separable, positive measure of fit. In the above, the $\beta-$divergence loss function is used for minimization~\citep{NMF_betaDivergenc}. $\beta$ is the single shape parameter used in the minimization and we employ $\beta = 2$ or squared Euclidean distance given by 
\begin{equation}
    D_2(\mathbf{X},\mathbf{X}') = \frac{1}{2} \sum_{i=1}^{m} \sum_{j=1}^{n} (\mathbf{X}_{i,j} - \mathbf{X}'_{i,j})^2 
\end{equation}
An alternating non-negative least squares (ANLS) algorithm is used for minimization that operates by alternately minimizing $D_\beta(\mathbf{X},\mathbf{WH})$ with respect to $\mathbf{W}$ and $\mathbf{H}$ such that all matrices are constrained to be non-negative. The above algorithm has been implemented with PyTorch by Maffetone et. al. ~\citep{cNMF} which is different from a canonically implemented NMF~\citep{McAuliffe}.

It is noted that the canonical NMF has some drawbacks~\citep{cNMF} and from the standpoint of scientific data analysis, a major one is that the output components may not correspond to physically-interpretable entities. 
The above drawback is circumvented by using cNMF where the initial values for $\mathbf{W}$, $\mathbf{H}$  -  $\mathbf{W}_0$ and $\mathbf{H}_0$ respectively, are constrained using prior knowledge and preset before the minimization as a constraint. Therefore, either the weights $\mathbf{W}_0$ or components $\mathbf{H}_0$ can be constrained at a time for factorization. In the following work, we constrain the components $\mathbf{H}_0$ and solve for the corresponding weights. In the next section we describe how this mathematical framework is implemented for EBSP data.

\subsection{Data Processing}
\label{Methods:Data_Processing}

\begin{figure}[htp!]
    \centering
    \includegraphics[scale = 0.06]{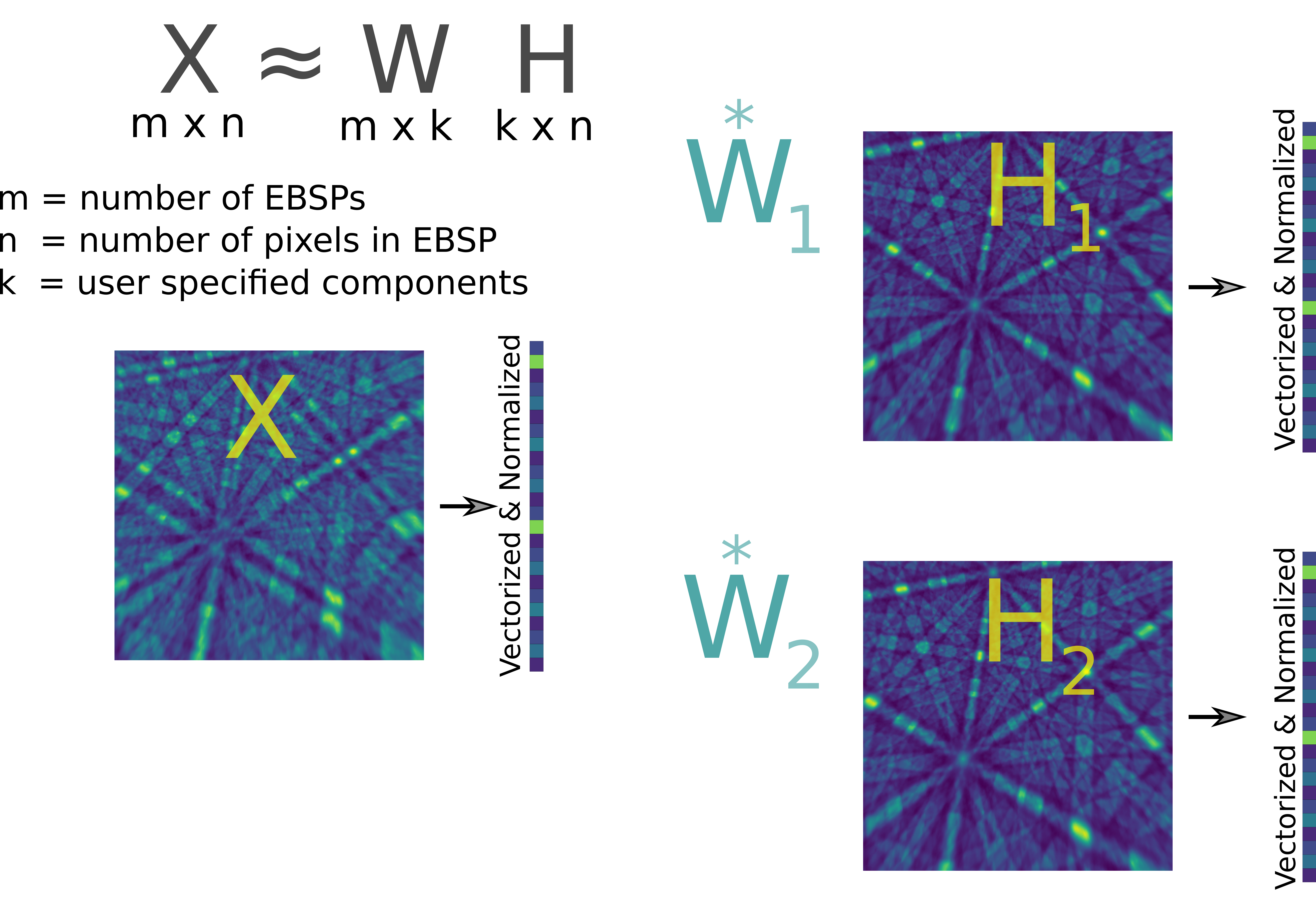}
    \caption{Schematic of the matrix algebra used in constrained non-negative matrix factorization. The data matrix $X$, the constrained components $H$ (preset EBSPs) and the unknown weights $W$.}
    \label{fig:cnmf_schematic}
\end{figure}
Each raw EBSP is first subtracted by a static background and a dynamic background but not processed further. Subsequently, the intensities for each pattern are normalized in the range 0 to 1 and vectorized as a single continuous vector.
For the analysis of experimental data, the chosen region of interest is shown in Figure~\ref{fig:cNMF_results}. Component EBSPs are selected from grain interiors marked by red markers in Figure~\ref{fig:cNMF_results}. To obtain sharp component EBSPs from the marked regions, a 3x3 pixel area is selected and averaged using a Gaussian 3x3 kernel. These component patterns $C_{1-6}$ are also shown in Figure~\ref{fig:cNMF_results}.

The component patterns $C_{1-6}$ shown in Figure~\ref{fig:cNMF_results} are very close with respect to their orientation.
This can be rationalized based on the fact that the chosen microstructure is essentially a single crystal with only small amounts of misorientations ($< 0.1\degree$), hence the patterns $C_{1-6}$ \textit{must} look very similar.
Conventional Hough-based EBSD methods are unsuitable to analyze such a microstructure and, therefore, we employ the Rotation Vector Base Line (RVB)-EBSD method~\citep{RVB_EBSD} for a baseline orientation mapping as reference.
The RVB-EBSD method is an HR-EBSD technique specifically designed to study crystal mosaicity and fine microstructural details.
Such microstructures are associated with slightly deviating growth paths of dendrites during directional solidification of Ni-based SX superalloys in a Bridgeman process~\citep{Hallensleben}.
RVB-EBSD utilizes a normalized cross-correlation function along with a rotational vector baseline function to determine pattern shifts caused by changes in orientation relative to a selected reference pattern.
This allows to detect small angular deviations ($< 0.1\degree$) clearly.
Therefore, consistently in this paper, we use the RVB-EBSD analyzed microstructure as a reference benchmark to compare our results obtained using cNMF.
For example in Figure~\ref{fig:cNMF_results}, the central figure (orientation map) and Figure~\ref{fig:KAMcomparison} (Kernel Average Misorientation).
Additional details of the RVB-EBSD method and the applied color coding schemes can be found in the relevant publication~\citep{RVB_EBSD}.

To prepare data for cNMF, we first take note that each EBSP with a resolution of $\mathrm{n}_{\mathrm{px}} \times \mathrm{n}_{\mathrm{px}}$, is an image with gray values from 0-255 and therefore by construction non-negative.
As described in Equation~\ref{eq:equation1}, $\mathbf{X}$ forms a matrix with dimensions $m \times n$, where $m$ is the number of independent measurements or scanned points and $n$ is the feature dimension of an EBSP or total pixels $N = \mathrm{n}_{\mathrm{px}} \times \mathrm{n}_{\mathrm{px}}$.
To analyze a grain boundary region between two grains, cNMF is carried out using only 2 known component patterns - $\mathbf{H_1}$, $\mathbf{H_2}$. These component patterns are selected from the $C_{1-6}$ shown in Figure \ref{fig:cNMF_results}, depending on the analyzed region of interest.
The constrained factorization then returns a pair of weights - $\mathbf{W}$ corresponding to each component $\mathbf{H_1}$, $\mathbf{H_2}$.
These 1-dimensional weights are then reshaped into respective 2-dimensional spatial dimensions for visualization.
In this way we obtain 2 weight maps - $w_1$,$w_2$, corresponding to each constrained component.

\section{Results \& Discussion}

\subsection{Mixing Simulated Patterns}
\label{Results:MixingSimulatedPatterns}

As a first step, we test the theoretical angular/factorization resolution limits of the method using artificially generated overlapping patterns through intensity mixing~\citep{Brewer_Multivariate}. This test can be interpreted as a simulation of the pattern overlap conditions near a grain boundary~\citep{Overlap_Tong}. 
An arbitrary orientation is chosen and a Kikuchi pattern is extracted from a gnomonic projected space of a dynamically simulated master pattern~\citep{Callahan_DI}.
Similarly, another pattern is extracted with a relative misorientation angle $\omega$ with respect to the first pattern. These two patterns are linearly mixed into 20 intensity fractions, which includes the two original \textit{pure} patterns as the limiting cases.
Further data processing is the same as described in Section \ref{Methods:Data_Processing}. Note however, that in this case the predicted weights are 1-dimensional, which can be compared to a linear behavior for verification. 

\begin{figure}[htp!]
    \centering
    \includegraphics[scale = 0.12]{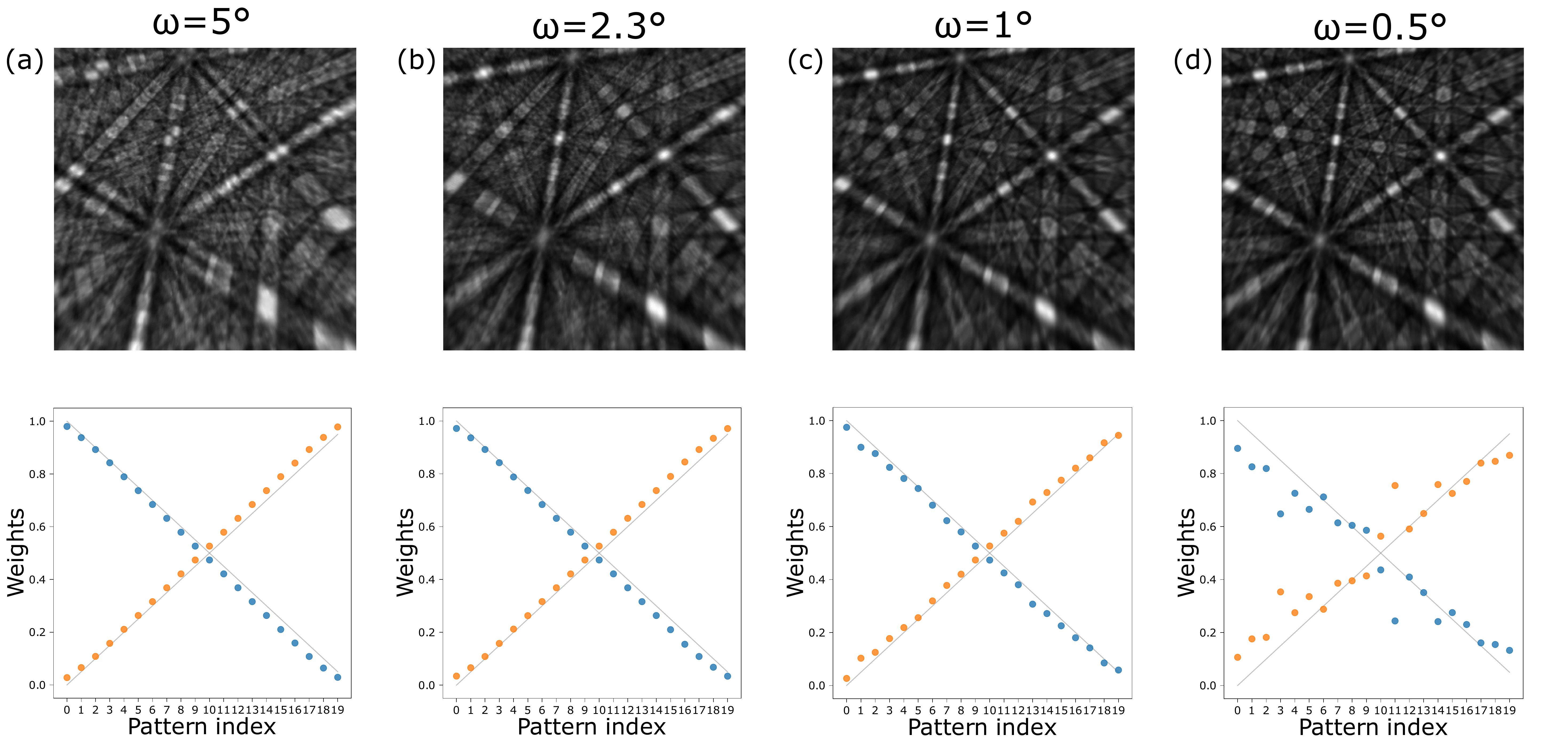}
    \caption{Constrained non-negative matrix factorization of a synthetic dataset showing predicted weights (points: blue = $w_1$, orange = $w_2$), compared to linear behavior (grey lines). The misorientation angle  between the component pattern used for synthetic EBSP generation is (a) $\omega = 5\degree $ (b) $\omega = 2.3\degree $ (c) $\omega = 1\degree $ (d) $\omega = 0.5\degree $. The Kikuchi patterns in each subfigure show the 10th pattern in the index that has equal ratio of mixing intensities.}
    \label{fig:Simulated_cnmf}
\end{figure}

Figure~\ref{fig:Simulated_cnmf} shows the resulting weights of the factorization of patterns oriented along $[110]$ direction and misoriented by $\omega$. We consider 4 cases in which $\omega$ is gradually decreased as $\omega=5\degree,\,2.3\degree,\,1\degree,\,0.5\degree$.
The displayed overlapping patterns in Figure~\ref{fig:Simulated_cnmf} correspond to the 10th index where there the original patterns are mixed with equal weight intensities.  
Figure \ref{fig:Simulated_cnmf} (a) shows a superimposed Kikuchi pattern with visible overlap of two patterns, a common case in grain boundary regions. The predicted weights $w_1$ and $w_2$ for different mixing ratios follow the grey lines or the linear interpolation in the same manner as they are mixed.
Note that the pattern overlap is clearly discernible in the image, due to a relatively large misorientation angle $\omega=5\degree$.
Besides the misorientation, the result also highlights that very small changes in pattern intensities or weights are captured using cNMF as 20 different weights, and therefore mixing ratios, are successfully resolved. This ability to resolve mixing ratios may be called as \textit{factorization resolution}, which is akin to crossing over a grain boundary in discrete steps or $\omega/$steps. Herein, the number of steps remain the same.
In addition, each data point is independently predicted in cNMF. In other words, each EBSP is individually factorized, without any influence of the outcome of other data points.

With a decrease of $\omega$ to $2.3\degree$ in Figure \ref{fig:Simulated_cnmf} (b), it becomes difficult to visibly discern the overlap of two patterns. Instead, only a blur is visible, e.g. in the bottom left corner.  Nevertheless, cNMF successfully predicts the calculated weights as per the linear mixing.

Further decrease of $\omega$ to $1\degree$, shown in Figure \ref{fig:Simulated_cnmf} (c), leads to some data points which begin to deviate from linearity.
Even though there is hardly any visible overlapping signature in the Kikuchi pattern itself, the cNMF method performs well even in this case.

A misorientation of $\omega = 0.5\degree$, as shown in Figure \ref{fig:Simulated_cnmf} (d) constitutes a case beyond the factorization limit.
At this misorientation angle and factorization limit of $\frac{0.5}{20}=0.025$, the weights cannot be determined accurately any more. 
Note however, that here the misorientation $\omega = 0.5\degree$ is not necessarily the angular accuracy limit of the method. Instead, it is only the factorization limit for the present case. To attest to this, notice that even though the predicted weights deviate from absolute linearity, the data still follows a linear trend. As we shall see in the later section using experimental patterns, the angular resolution limit can be lower. In general, increasing the factorization limit is possible by either increasing the misorientation angle or decreasing the number of mixed patterns. The latter is a scenario akin to which is a scenario akin to decreasing the number of steps when moving across the grain boundary. 


Using the above analysis we are able to simultaneously test two parameters.
The first is the sensitivity of cNMF to misorientation and the second is sensitivity to mixing steps.
In practice the first is akin to resolving small misorientations between grains while the second is related to step size.
We demonstrate that the cNMF method successfully predicts these variations.
Moreover, we conclude that it is possible to locate grain boundaries in a microstructure using a weight based metric.

\subsection{Experimental Patterns}
\label{Results:ExperimentalPatterns}

\begin{figure}[htp!]
    \centering
    \includegraphics[scale=0.25]{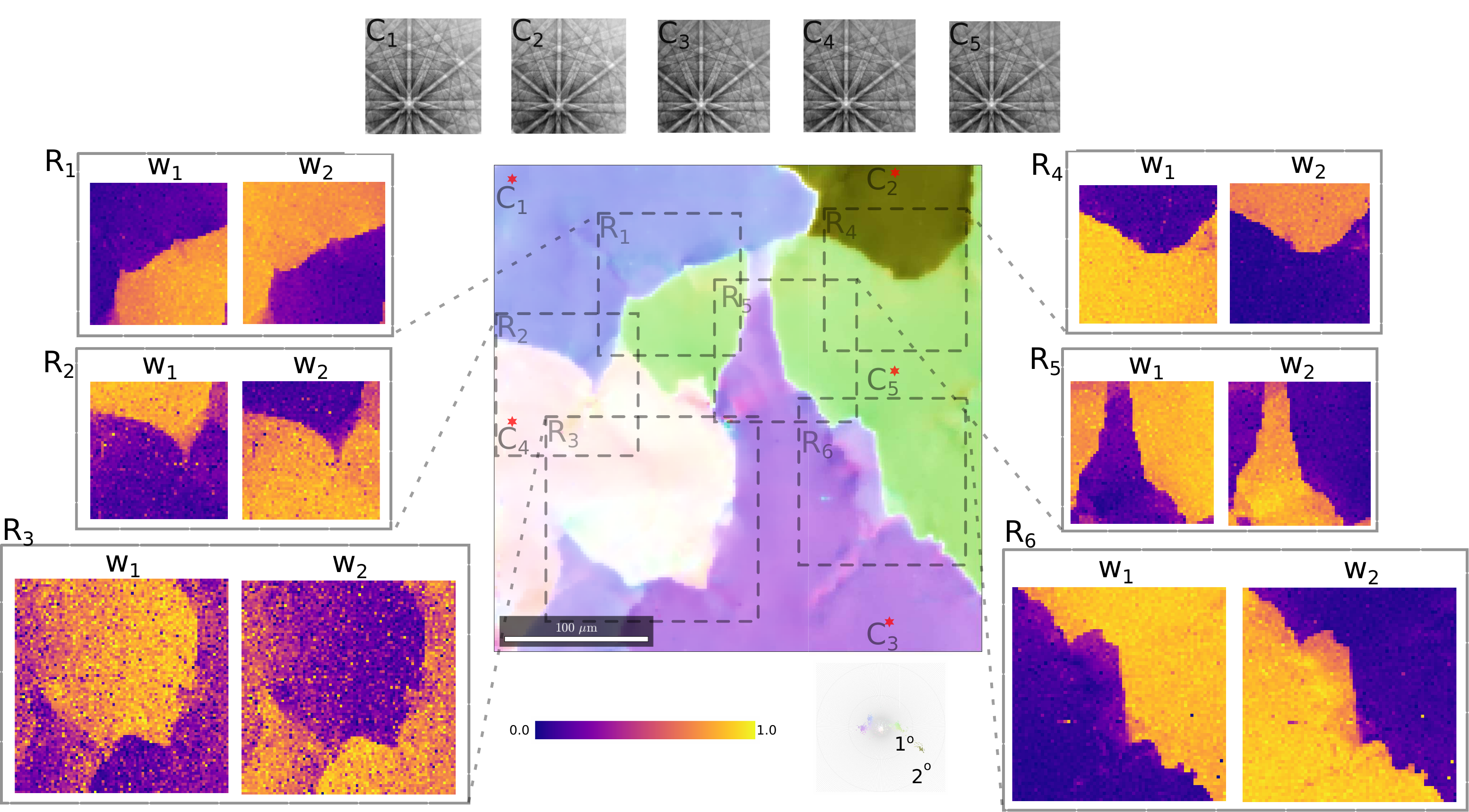}
    \caption{Weight maps of cNMF factorization chosen from different regions of the scanned area. The central figure shows the full scanned area and is colored according to a pole figure generated using RVB-EBSD method with the pole along a $[001]$ direction as a $2\degree$ pole. Within the pole figure, specific regions are highlighted using  \textcolor{red}{red} stars. They mark the locations chosen for the constrained components. Note that there are 5 locations of component patterns in total corresponding to 5 separate dendritic regions. All weight maps show a low-angle grain boundary between two regions and, correspondingly, two weight maps which show individual variations in weight $w_1$ and $w_2$ of the chosen patterns as constraints. }
    \label{fig:cNMF_results}
\end{figure}

The use of simulated patterns is an essential step to test the limits of the method but it is difficult to simulate real patterns containing noise, background signal, beamshifts, etc., and which are affected by sample surface quality, sample preparation artifacts, plastic deformation zones, surface relief and contamination issues.
Therefore, in this section we apply cNMF on an experimental dataset of an ERBO1/C Ni-based single crystal superalloy specimen.
The details of specimen preparation are provided in Section \ref{Methods:Experimental}.
This dataset is specifically selected due to the prevalence of low angle grain boundaries (the grains are within $1.5\degree$ orientation spread).
This approach further tests the practical limits of the cNMF method by resolving patterns that are very similar (also see $C_{1-6}$ in Figure \ref{fig:cNMF_results}).
We compare our results to a RVB-EBSD based high-resolution analysis~\citep{RVB_EBSD}, which is state-of-the-art in analyzing such microstructures.

Figure~\ref{fig:cNMF_results} (R$_{1-6}$) shows the reconstructed weight maps using cNMF.
The analysis is made using small sub-sections of the scanned area, such that each region of interest contains a low-angle grain boundary between two dendritic regions.
The constrained components for each region are extracted from within the grain interior marked C$_{1-6}$ with red stars. For example, in region R$_{1}$ the constrained component EBSPs are C$_1$ and C$_5$. Similarly for region R$_{5}$ they are  C$_3$ and C$_5$. In this manner, corresponding to each component combination, a weight map is obtained, shown using $w_1$ and $w_2$. Note that the weight maps are shown using both $w_1$ and $w_2$ as both together can be used as indicators to demonstrate variations in each individual weight.
Unlike the synthetic pattern analysis consisting of superimposed patterns in section~\ref{Results:MixingSimulatedPatterns}, in this case only the grain boundary EBSPs are expected to exhibit pattern overlaps whereas a single pattern (orientation) typically dominates in the grain interior. This can be seen in Figure~\ref{fig:cNMF_results} region R$_1$, where each weight map $w_1$, $w_2$ has distinct bright and dark colored regions which invert symmetrically across the grain boundary. That is, weights are inversely proportional which allows us to demarcate the extent of each grain. Notice that in terms of misorientation, each pair of two grains lie within $1\degree$ pole as shown in the RVB-EBSD pole figure in the center bottom. In addition, also note that in the pole figure the grain boundary pixels may not be indexed to distinct orientations because of pattern overlap (the grain boundary pixels are white), which is not the case with weight maps $w_1$, $w_2$. Thus, using weights, the extent of the grain boundary is distinctly known. Furthermore, in the weight map $w_1$ the intensity of the pixels varies within the grain which is indicative of small orientation changes, consistent with the RVB-EBSD pole figure.

In region R$_4$, the grain boundary is clearly visible in the weight maps. Interdentritic orientations, however, are not visible, consistent with RVB-EBSD pole figure.

In region R$_5$, two boundaries exist between intersecting dendritic regions which consist of only 2 grains. The corresponding weight maps exhibit similar pixel intensities that lie within the same grain which distinctly demarcates the grain boundaries. Moreover, weight map $w_2$ shows a variation in intensity within the respective grain (bottom left), that is consistent with interdendritic orientation changes in the pole figure.

In region R$_6$, weight maps demarcate the extent of the grain boundary as well as local changes in orientation along the grain boundary in map $w_2$. The above cases demonstrate that cNMF is able to detect orientation variations which are on par with the RVB-EBSD reference (orientation changes are less than $0.25\degree$). Note however, the variations in intensities in region R$_6 $ within a dendritic region implicitly arise due to sensitivity of the method as the factorized EBSPs are dissimilar to the component EBSPs. This implies that deviations from the component pattern lead to changes in weights, even though they arise from pattern degradation and not alone from orientation changes. Following this, it can also be argued that all EBSPs that deviate from the two chosen component EBSPs may correspond to a 3rd orientation or grain.

This aspect is brought out more clearly in region R$_2$ where, apart from a grain boundary, a triple junction exists in the top right corner. In the weight maps $w_1$, $w_2$, two dominant grains are distinctly visible. Herein, the constrained components are C$_1$, C$_4$. Yet, the predicted weights in the 3rd intersecting dendritic region (right corner), which is not used as component in cNMF, are very similar in intensities for both $w_1$ and $w_2$ maps.
Thus, attesting to the argument that any orientation that is not used in cNMF may still manifest itself in the form of weight intensity variation. This, generally, allows the segmentation of more than two grains even if only two components are used, albeit with some loss of accuracy. More discussion on this aspect is provided within Section \ref{Results:General Discussion}.

In Region R$_3$, the component EBSPs are located at C$_3$ and C$_4$, where there is a change in orientation within the left grain. In other words, the grain boundary region is slightly misoriented to the orientation of the  C$_4$ location. Furthermore, the misorientation angle between the two grains is lowest in this region ($\approx0.5\degree$) as shown in the central pole figure. Consequently, the calculated weight maps exhibit a considerable amount of noise even through the demarcation of dendritic and interdendritic grain boundaries are successfully identified. Thus, this case forms the limiting case of the cNMF method. Moreover, this demonstrates that the weight maps are sensitive to the location of the selected component patterns in the scan space with respect to the region of interest.

\subsection{Resolving the grain boundaries}
\label{Results:ResolvingGrainBoundaries}
\begin{figure}[htp!]
    \centering
    \includegraphics[scale = 0.5]{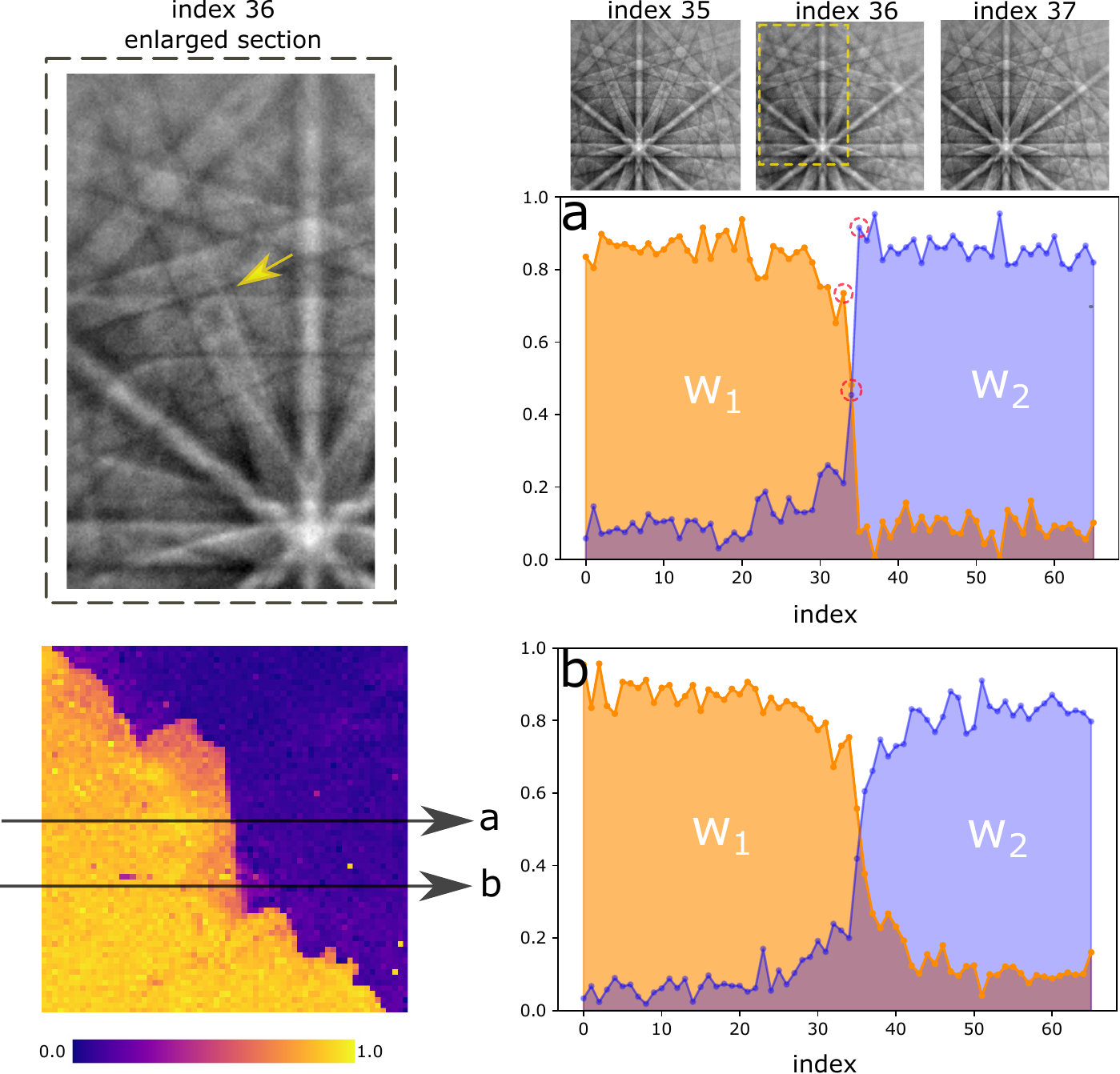}
    \caption{Profiles of individual weights $w_1$, $w_2$ across two grain boundaries: (a) sharp boundary and (b) smeared boundary. The weight map $w_1$  of region R$_6$ is used to depict the microstructure. Along profile `a', the EBSPs at 3 points at the grain boundary from index 35-37 are shown. The location of these EBSPs along the profile are indicated with red circles in (a). The enlarged section of the EBSP at index 36 is shown where a faint signature of pattern overlap is visible and indicated with a yellow arrow.}
    \label{fig:weight_profile}
\end{figure}

Using a weight based metric instead of orientation allows us to resolve grain boundaries by leveraging the change in weights and, thereby, address the issue of pattern overlaps. Consider Figure~\ref{fig:weight_profile}, which shows the weight profiles across a grain boundary for two cases -  (a) across where orientation appears to transition sharply from one pixel to the other and (b) where the grain boundary appears \textit{smeared} out.
Along profile (a), weights $w_1$ and $w_2$ exhibit a discrete step-like change across the grain boundary. Notice the relative symmetric reversal of the weights across the grain boundary. This implies that the grain boundary is sharp and patterns do not have a measurable overlap within the grain interior but only at the interface. To better explore this behavior we consider three distinct patterns which lie along the grain boundary profile (a). Pattern indices 35,36,37 lie across profile `a' at the transition from left grain to the right.
Their respective weights are marked with red circles in the Figure~\ref{fig:weight_profile}(a).
Pattern 36 exhibits an overlap of patterns 35 and 37  which is, however, very hard to visually discern.  An enlarged section of pattern 36 shows a faint secondary band marked by a yellow arrow. This indicates that there is some pattern overlap of patterns 35 and 37 which is successfully resolved. The above results thus depict a use case which is comparable to the scenario tested using simulated patterns in Section~\ref{Results:MixingSimulatedPatterns}.

On the other hand, along profile `b', weights $w_1$ and $w_2$ exhibit a gradual change across the grain boundary. Across the grain boundary along profile `b', the weight map shows a visibly smeared grain boundary which indicates that the orientation changes gradually. Nevertheless, the position of the grain boundary can be estimated at the point where the two weights $w_1$, $w_2$ make a transition. This scenario is comparable to the scenario where the grain boundary is crossed in several steps as discussed previously (Sec.~\ref{Results:MixingSimulatedPatterns}). Herein, the relative transition of weights $w_1$, $w_2$ can also be exploited to estimate the location of the grain boundary.

\begin{figure}[htp!]
    \centering
    \includegraphics[scale = 0.28]{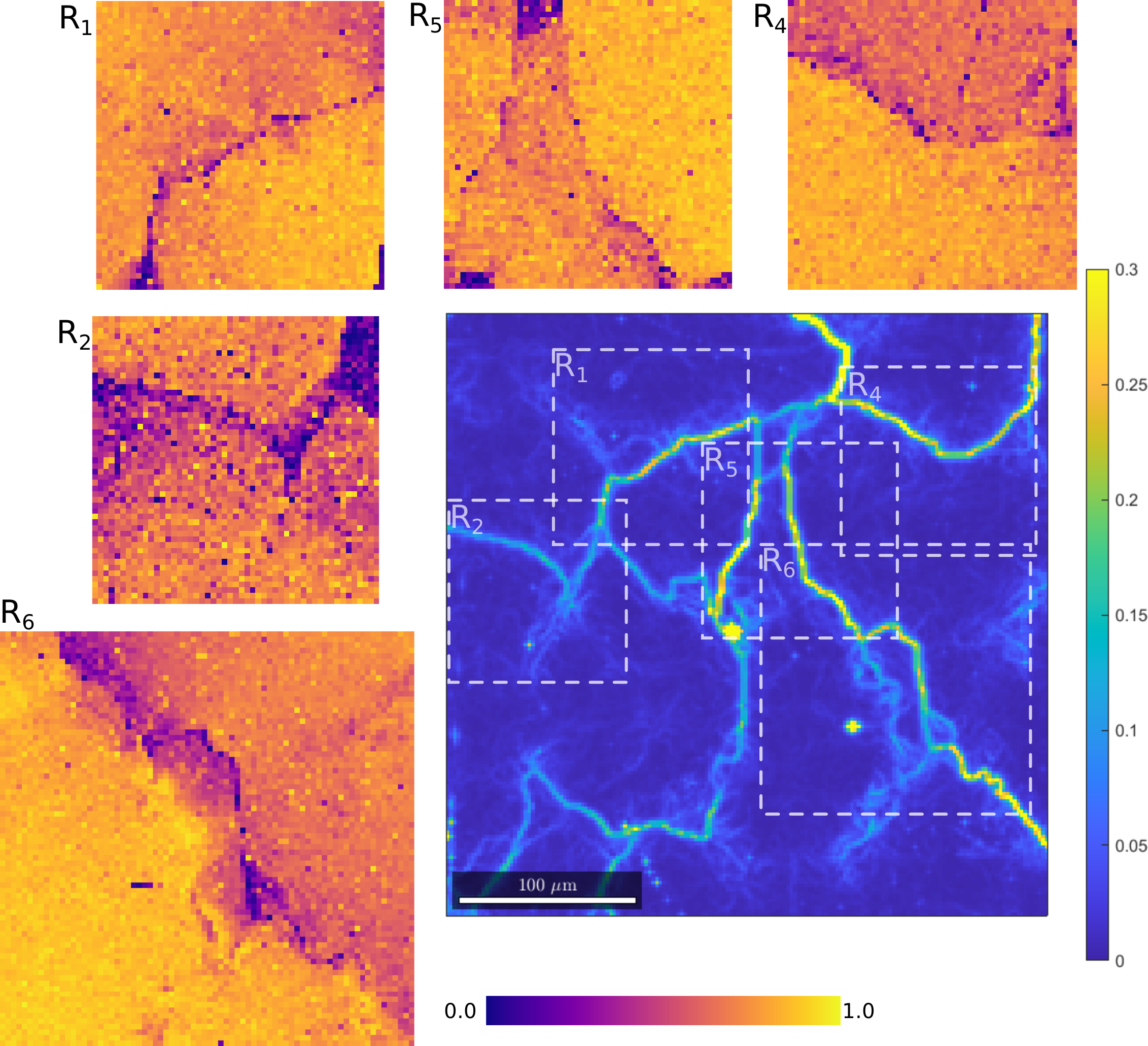}
    \caption{Absolute weight difference maps of different regions and comparison to a KAM map. The KAM map is obtained after processing the pole figure map in Figure \ref{fig:cNMF_results}.
    Enlarged versions of the regions are shown in their correct relative size. The highlighted regions are the same as selected in Figure \ref{fig:cNMF_results}.} 
    \label{fig:KAMcomparison}
\end{figure}

To locate the grain boundary, we need to extract the information contained implicitly in the weight metric. The transition between very similarly oriented regions can be captured by calculating the absolute differences in weights $abs|w_1 - w_2|$ per pixel since pixels at the grain boundary have similar or equal weights as a consequence of pattern mixing.
Consequently, the absolute difference in weight is lower at and near grain boundary region compared to the grain interior (also consider the weight profile in Figure \ref{fig:weight_profile} for an intuition). 
To test this hypothesis, we show the absolute weight difference maps of separate regions in Figure~\ref{fig:KAMcomparison}.
For comparison, we additionally present a Kernel Average Misorientation (KAM) map for the same region as obtained using RVB-EBSD~\citep{RVB_EBSD}.

In Figure \ref{fig:KAMcomparison}, region R$_1$, the grain boundary is revealed as dark colored pixels, where as the brighter pixels belong to the grain interior.
This follows from the hypothesis that the absolute difference in weights in the darker pixels should be lowest while the grain interiors would show relatively larger values.
A comparison with the KAM sub-region shows that in the difference weight map the grain boundary is much sharper and well defined. Other interesting features are the apparent discontinuities in dark pixels at the grain boundary which may be understood as a consequence of the step size and orientation of the grain boundary with respect to the scan direction~\citep{Overlap_Tripathi}.
Darker pixels are clustered around regions where the grain boundary curvature changes.
Local orientation changes arising from sub-grain boundaries are also visible in these maps as variations in intensity.
Overall, it is possible to resolve a grain boundary at the level of a pixel.
This resolution would not be possible using a KAM analysis as each pixel is assigned to a particular orientation which then leads to an average misorientation with respect to the surrounding pixels.
In other words, depending upon the size of the kernel, a grain boundary will always be a few pixels wide and also continuously connected.

Our proposed approach is capable of detecting regions of pattern overlaps very effectively and efficiently. In cases where minimal overlap exist, we can nevertheless capture the relative variations from one grain to the other.
For example, in Figure~\ref{fig:KAMcomparison}, region R$_5$, the grain boundary can be identified as different pixel intensities in the two grains.
This allows us to sharply demarcate the grain boundary.
A similar scenario unfolds in Figure~\ref{fig:KAMcomparison}, region R$_4$, where the grain extent is distinctly demarcated as a change in mean intensity, along with some darker pixels at the grain boundary. That is, the left grain is brighter than the right grain and separated by a dark grain boundary region.

Even maps derived from noisy weight maps, such as Figure~\ref{fig:KAMcomparison}, region R$_2$, allows the detection of the grain boundary region very accurately.
Another interesting feature is notable in this region.
The small portion of a third grain (top right) manifests in the form of dark pixels. Understandably so, because the absolute difference scheme entails low intensity pixels for that portion of the map, which belongs to a 3rd grain orientation which is not part of the constraints.
This, in principle, allows to resolve triple junctions around this region similar to the predictions of Brewer~\citep{Brewer_Multivariate}.
However, the analysis of triple junctions does not lie within the scope of the present work but can be subject of future studies.

Finally, Figure~\ref{fig:KAMcomparison}, region R$_6$, reveals that a combination of the previously described scenarios can be present.
The grain boundary is sharp and denoted by dark pixels is specific regions.
A relative variation of mean intensity exists between the two adjoining grains, thereby allowing us to demarcate the extent of each grain.
That is, the left grain is brighter than the right grain.
In addition, some areas around the grain boundary show substantially darker and smeared-out transitions at the grain boundary.
This can be rationalized as a combined affect of the presence of sub-grain structures and because the component EBSPs are slightly different from the sub-grain structure EBSPs. These features, however, only show up as gradual variations in intensity. 

\subsection{General discussion}
\label{Results:General Discussion}

\begin{figure}[htp!]
    \centering
    \includegraphics[scale = 0.22]{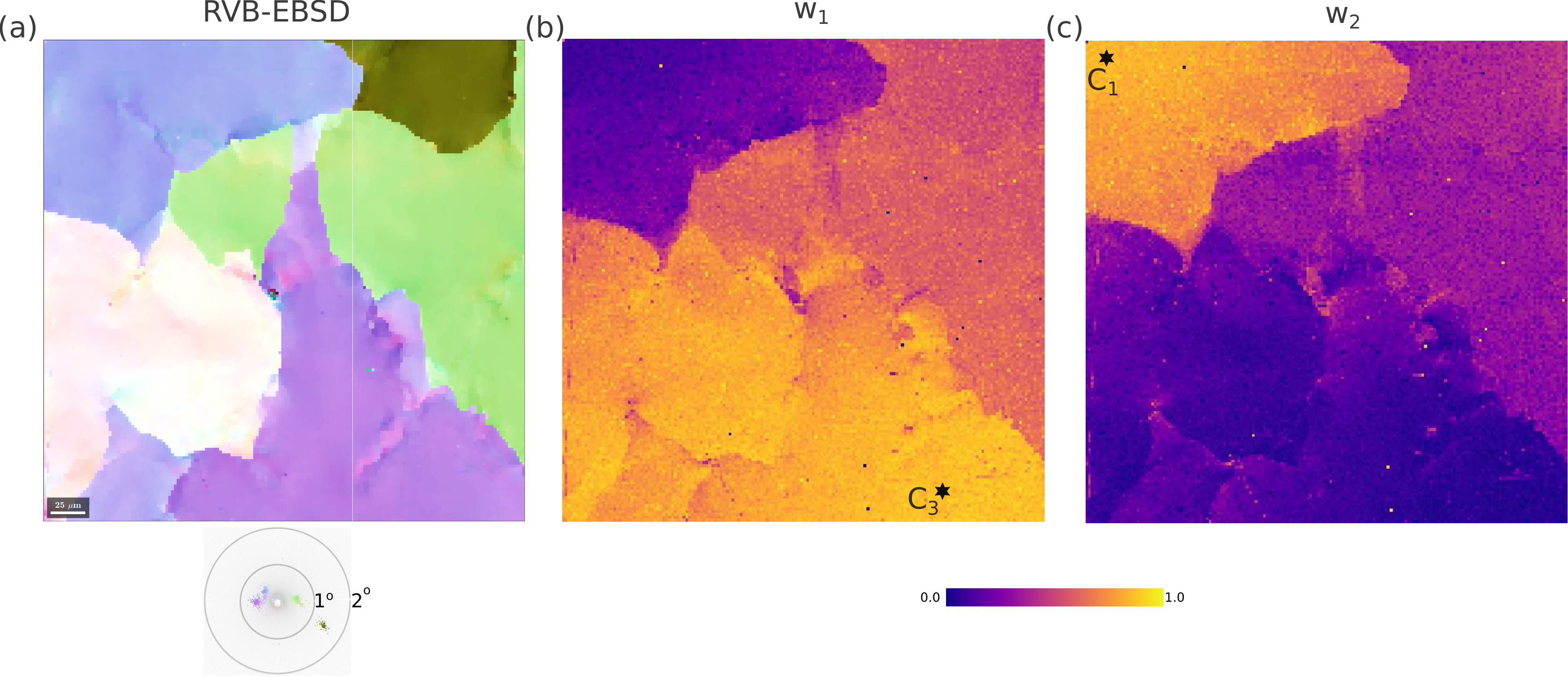}
    \caption{A1: cNMF factorization of the entire region of interest using weight-maps $w_1$ - (b) and $w_2$ - (c). The component EBSPs originate from C$_1$ and C$_3$ which have also been marked. For comparison an RVB-EBSD generated pole figure has also been provided - (a)}
    \label{fig:sup1}
\end{figure}

\begin{figure}[htp!]
    \centering
    \includegraphics[scale = 0.12]{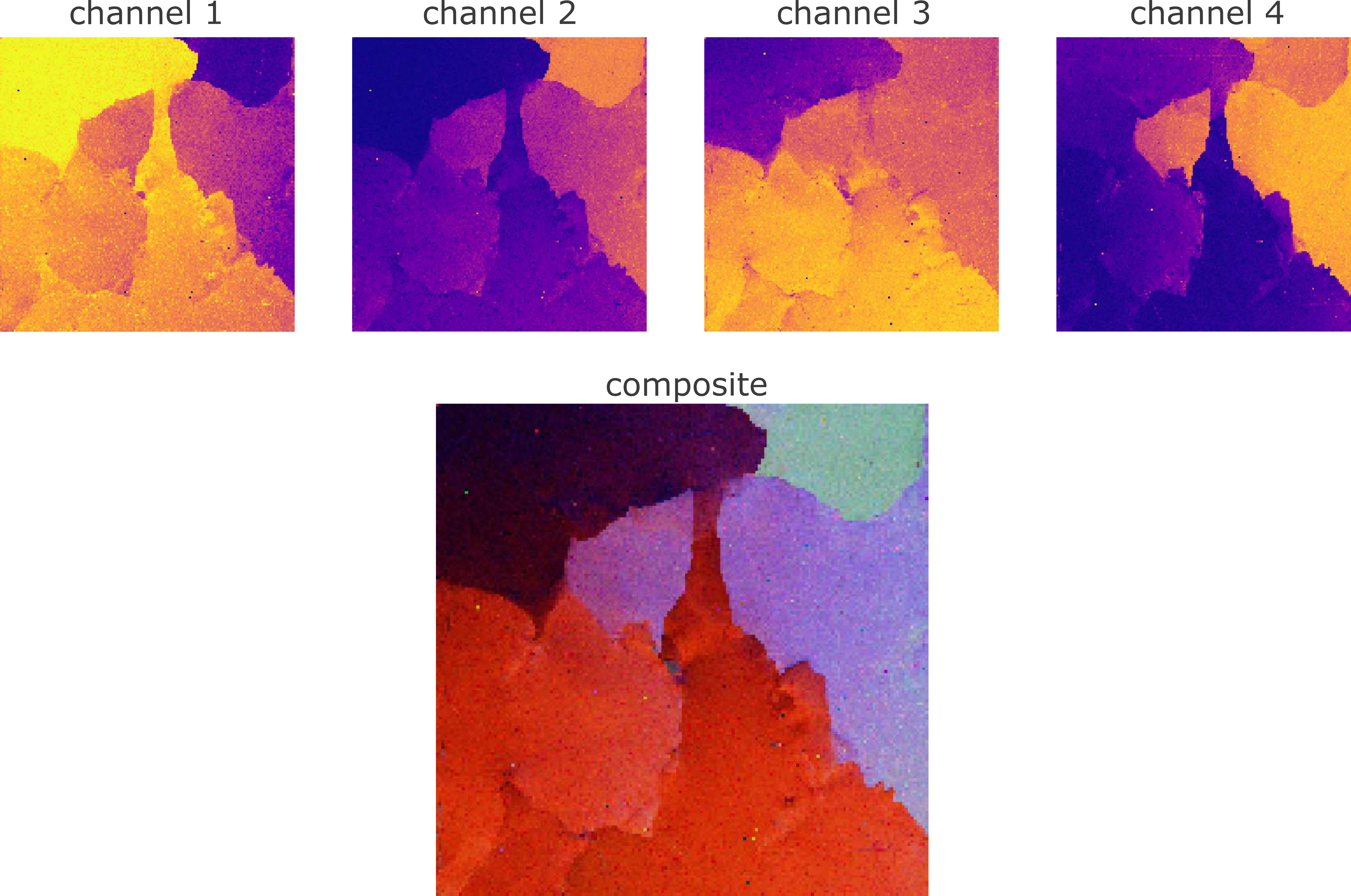}
    \caption{A2: Constitution of a 4 channel composite color image formed by combining 4 different weight-maps of the entire region of interest.}
    \label{fig:sup2}
\end{figure}

First and foremost, the cNMF method provides a novel method of segmentation as compared to an orientation indexing approach.
The use of a weight metric to resolve the extent of each grain allows us to capture orientation changes within the grain as well, which is on par with HR-EBSD methods.
While the weight maps shown in Figure \ref{fig:cNMF_results} belong to a grain boundary between only two grains, it is in principle also possible to resolve larger areas.
For example in Figure \ref{fig:sup1}, a two-component segmentation with one weight map is enough to demarcate the extent of dendrites and interdendritic features. In the figure $w_1$ map shows the intensity of weights which are most similar to the component EBSP at C$_3$. Notice the level of detail within the large dendrite that can be captured along with a clear demarcation of grain boundaries. 
To further extend the same idea, we can also calculate multiple two-component factorizations and then combine the results together. For example in Figure \ref{fig:sup2} four different instances of factorization are shown with component EBSPs originating from the grain interiors. These four weight-maps are then combined to form a 4 channel color composite image that shows the microstructure in fine detail. Herein, similarity in colors signify similarity of orientation as well.

Note that this is possible in our experimental data set because the patterns belong to similarly-oriented grains or very similar EBSPs.
In case of large angle grain boundaries, the regions which deviate substantially from the grain of interest exhibiting low intensities.
This may be construed as a drawback of the method. However, we propose our method as a companion tool useful in special cases such as those presented above or in segmentation problems where indexing crystallographically similar phases creates its own problems~\citep{McAuliffe,Overlap_Lenthe}.

The second key aspect of using cNMF is the ability to factorize overlapping pattern into their constituents which follows a paradigm of employing a weight-based metric.
This is advantageous as the binary classification of a pixel that is used in orientation imaging~\citep{Overlap_Tong} is avoided. So far, the method of reducing the resolution of the scan step is used to accurately resolving grain boundaries~\citep{Overlap_Zaefferer}.
However, a weight-driven approach can allow to clearly locate a grain boundary without changing the resolution limits (accelerating voltage) of the electron beam itself. A continuous variable in the form of weights allows us to appreciate the smooth transitions across a grain boundary and detect EBSPs with pattern overlaps distinctly.
These transitions may arise either due to inherent orientation changes or due to the step size during measurement. 
Following the analysis in Sections~\ref{Results:MixingSimulatedPatterns} and~\ref{Results:ResolvingGrainBoundaries}, we can also argue that the grain boundary can be resolved to a single pixel, irrespective of the step size. That is, the most probable location of the grain boundary should be at pixels with largest amount of overlaps. This aspect of cNMF can be particularly useful in spatially resolving nanoscale phases where the electron-matter interaction volume is of comparable size to the characteristic length scale of the phase~\citep{Overlap_BroduAimo}. In which case resolving pattern overlaps and cleaning data is of utmost importance for proper characterization. On a similar node, the ability of the method to detect small intensity variations can be useful in pseudo-symmetry problems~\citep{Chirality}, where there is a minor difference between patterns~\citep{SamePattern}. Overall cNMF can prove to be a viable alternate for such scenarios.

Finally, a comment on the robustness and efficiency of cNMF.
The method is already in use for various scientific analysis~\citep{cNMF} and allows for enormous flexibility in the input data.
We have implemented the scheme for EBSD data analysis using EBSPs.
In practice, however, a variety of signals can be used, as the method is not affected by material/signal/dimensions and therefore can work for different crystallographic orientations and phases.
For example, locating the boundaries of  $\gamma$/$\gamma'$ phases in superalloys~\citep{Gamanov2023}.
In addition, our proposed scheme would also work for a hybrid dataset where energy dispersive X-ray spectrocopy (EDS) data can be combined with EBSP signal~\citep{McAluffi2}, i.e. use a multimodal representation of a material or part of it.
With respect to efficiency, since each pattern/signal can be read and factorized independently it is also possible to parallelize for GPUs. In terms of speed, the method is less computationally expensive than the HR-EBSD methods but like such methods, requires storing Kikuchi patterns. Some part of this storage space can be minimized by only analyzing small grain boundary regions on the fly and discarding after analysis. Furthermore, the pattern overlap detection of cNMF can still be used efficiently within the the standard indexing procedure, as an add-on. To close, we note that any existing dataset where Kikuchi patterns are recorded can be re-analyzed, since our proposed method does not depend on any specific measurement settings or crystallographic information. Consequently more information regarding phases, orientations and defects can be extracted from already existing datasets.

\section{Conclusions}

We implement a semi-supervised constrained non-negative factorization (cNMF) method to segment and demarcate grain boundaries.
Using linear superpositions of synthetic patterns, we demonstrate the angular sensitivity and factorization resolution of this technique.
The process simulates two scenarios simultaneously - (1) the sensitivity to misorientation between patterns (2) sensitivity to \textit{factorization resolution}. Our method performs well in resolving misorientations considerably $<1\degree$.
This proves that the method can work effectively, in particular for low-angle grain boundaries.
\par
To test the limiting case of grain boundary segmentation (LAGB), we demonstrate the use of the method on a dataset of a technical single crystal superalloy.
The specimen comprises of dendrites which deviate by small misorientation angles, $\omega<1\degree$.
Using cNMF, the grain boundary regions are segmented using \textit{weight maps}.
Inherent to the method is the paradigm of segmenting a microstructure in the form of weights w.r.t. to user-defined references instead of  absolute orientations.
The weight maps clearly segment the grains and show features which are on par with RVB-EBSD method - an HR-EBSD technique specifically designed for mapping small misorientations during dendritic growth.
Using the weight metric, we are able to consider overlaps in patterns and show how more information can be extracted from the data.
The regions of highest pattern overlaps are regions where the absolute difference in weights is lowest.
Using this metric, the grain boundary can be resolved even at a pixel level, albeit with some nuances, which are discussed in detail.  
\par
The cNMF method is versatile and may find use in applications such as grain and phase segmentation with similar EBSPs and small-scale phases with extensive overlapping EBSPs. Furthermore, due to its robustness, hybrid signals can be used which use a combination of both EBSP and EDS signals.
While the processing speed may still be slower than the standard indexing speeds today, the method can be useful when used in conjunction with orientation imaging methods or in localized regions. 

\section*{Data statement}
Kikuchi pattern data set can be downloaded from Zenodo 
(\url{https:doi.org/10.5281/zenodo.10705510} ). An up-to-data version of the implementation of cNMF is available from a github respository (\url{https://gitlab.ruhr-uni-bochum.de/icams-mids/kikuchi_cnmf.git}) .

\section*{Conflict of interests}
The authors declare no competing financial interests or personal relationships that could have influenced this work. 

\section*{Appendix}

\newpage
\bibliographystyle{abbrvnat}
\bibliography{cNMF}



\end{document}